\documentclass[fleqn,10pt]{wlscirep}
\title{Highly tunable time-reversal-invariant topological superconductivity in topological insulator thin films}

\author[1,2,*]{Fariborz Parhizgar}
\author[2,+]{Annica Black-Schaffer}
\affil[1]{School of Physics, Institute for Research in Fundamental Sciences (IPM), Tehran 19395-5531, Iran}
\affil[2]{Department of Physics and Astronomy, Uppsala University, P.O. Box 516, S-751 20 Uppsala, Sweden}

\affil[*]{fariborz.parhizgar@ipm.ir}

\affil[+]{annica.black-schaffer@physics.uu.se}

\begin{abstract}
We study time-reversal-invariant topological superconductivity in topological insulator (TI) thin films including both intra- and inter-surface pairing. We find a nontrivial topology for multiple different configurations. For intra-surface pairing a $\pi$-phase difference between the intra-surface pairing states is required. We show that in this case the resulting topological phase is highly tunable by both an applied electric field and varied chemical potential. For spin-singlet inter-surface pairing, a sign-changing tunnel coupling present in many TI thin films is needed, and again, the topology can be tuned by electric field or doping. Notably, we find that the required inter-surface pairing strength for achieving nontrivial topology can still be subdominant compared to the intra-surface pairing. Finally, for spin-triplet inter-surface pairing we prove that the superconducting state is always topological nontrivial. We show that thin films of Cu-doped Bi$_2$Se$_3$ will likely host such spin-triplet inter-surface pairing. Taken together, these results show that time-reversal-invariant topological superconductivity is common in superconducting TI thin films and that the topological phase and its Kramers pair of Majorana edge modes is highly tunable with an applied electric field and varied chemical potential.
\end{abstract}
\begin{document}

\flushbottom
\maketitle
\thispagestyle{empty}

\section*{Introduction}

Topology in condensed matter systems has received a massive amount of interest in recent years. Starting with the discovery of topological insulators (TIs),\cite{KaneMelePRL,BernevigScience, KonigScience, Hsieh08} where nontrivial topology of the bulk band structure creates surface states with a Dirac-spectrum crossing the bulk band gap, topology is now also applied widely in the discussion of properties of superconductors and semimetals.\cite{TI-review, QiZhangRMP, Wehling-2014, BansilRMP}
In topological superconductors the zero-energy surface states are particularly intriguing as they are Majorana modes, due to their equal electron and hole characteristics.\cite{Majorana-1937, Kitaev-2001, Beenakker, Leijnse} The non-Abelian statistics of Majorana zero-modes makes topological superconductors very promising for topological quantum computation.\cite{Ivanov-2001, Nayak-2008, Sarma15}

Many topological superconductors explicitly break time-reversal symmetry. This is usually achieved by magnetic field or impurities, which together with Rashba spin-orbit coupling and superconductivity can give nontrivial topology.\cite{Sau-2010, Lutchyn-2010, Oreg-2010, ABSLinder11SNS, Choy-2011, Mourik-2012, Nadjperge2013, Nadjperge2014} A close alternative is superconducting TI surfaces, where Majorana modes appear in vortex cores or at ferromagnetic interfaces, again requiring magnetic fields.\cite{Fu-2008, Fu-2009, ABSLinder11QSHI, Wang-2012, Xu-2015}
However, time-reversal-invariant superconductors can also be topologically nontrivial, where instead the surface spectrum consists of a Kramers pair of helical Majorana modes.\cite{Qi2009PRL} Despite the doubling of the Majorana modes, non-Abelian operations have still been demonstrated in these systems.\cite{Liu-2014} Since superconductivity is generally very sensitive to magnetic fields, time-reversal-invariant systems have a very natural advantage in terms of stability of the superconducting phase.

Proposals for time-reversal-invariant superconductors have mainly been based on multi-channel systems, since noninteracting single-channel systems with conventional $s$-wave superconductivity cannot be topological.\cite{Haim2016Arxiv}
For example, multiband/orbital systems and $s_{\pm}$-wave iron-based superconductors have been shown to be able to host a nontrivial phase.\cite{Zhang-spm,Deng-2012, Wang2014} Double wire systems with Rashba spin-orbit coupling have also been studied, where a $\pi$-phase difference of the superconducting order parameter between the two wires,\cite{Keselmman-2013} or a strong inter-wire pairing\cite{Flensberg-2014, Jelena-2014, Danon2015, Ebisu2015} have been shown to give nontrivial topology. Similar topological inter-wire superconductivity has been found using two 1D edges from two independent 2D TIs.\cite{Jelena-2014-2} A 1D extended Hubbard chain also has been shown to support topological superconductivity in the presence of both singlet and triplet pairing.\cite{Kuei-2014}
In addition, a $\pi$-phase difference has been shown to give a nontrivial topology in a simple model for unbiased TI thin films\cite{Liu&Trauzettel} and by introducing random magnetic moments between a superconductor and a TI surface.\cite{Constantin} Similar criteria have been found in double layer Rashba materials with strong interactions.\cite{Nakosai-2012}
However, all these systems have shown zero or very limited tunability of the topological phase by experimentally controllable parameters, which both limit applications and possibilities to detect nontrivial topological phases. 

In this work we study conventional $s$-wave superconductivity in realistic models of superconductivity in three-dimensional (3D) TI thin films. For TI films thinner than 6 quintuple layers (QL), an energy gap opens in the surface states due to inter-surface tunneling.\cite{Zhang-2010} However, sufficient doping produces a finite density of states at the Fermi level, such that thin films can easily enter a superconducting phase either by proximity to external superconductors,\cite{Wang-2012, Xu-2014} or possibly by an intrinsic mechanism.\cite{intrinsic-SC1,intrinsic-SC2,intrinsic-SC3,intrinsic-SC4,intrinsic-SC5,intrinsic-SC6,intrinsic-SC7} To allow for the most general situation we consider both $s$-wave intra-surface pairing and inter-surface pairing, since inter-surface pairing has been shown to be present in proximity-induced superconducting thin films through crossed Andreev reflection\cite{CAR1, CAR2} and from a direct proximity effect.\cite{Parhizgar14} 
We show that intra-surface pairing can lead to a topologically nontrivial state when there is a $\pi$-phase difference between the two surface pairing terms. Further, spin-triplet inter-surface pairing is odd in surface index, which we show always result in a topologically nontrivial state. For nontrivial topology with spin-singlet inter-surface pairing, the thin film tunneling needs to change sign with momentum, which is very common in TI thin films.
Most importantly, in all cases the topological phase is highly tunable with an applied electric field or by varying the chemical potential. This allows for direct tuning of time-reversal-invariant topological superconductivity and its associated Kramers pair of Majorana edge modes in TI thin films.  
With proximity-induced superconductivity already established in 3D TI thin films and intrinsic superconductivity found in several TIs, our results should have direct experimental relevance.

% ----------------------------------------- %
% MODEL:
% ----------------------------------------- %

\subsection*{Model Hamiltonian}
\begin{figure}[th]
\centering
\includegraphics[width=14.0cm]{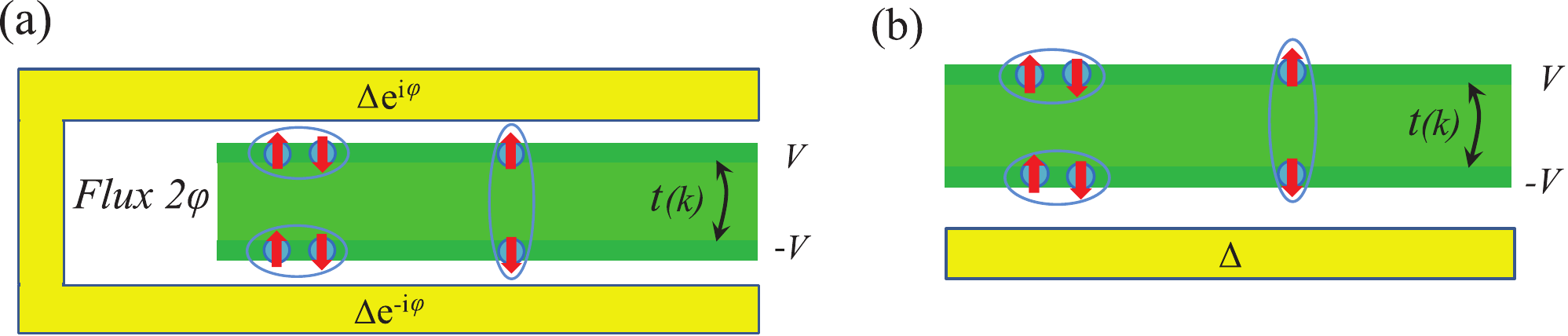}
\caption{Two different schematic setups for proximity-induced superconductivity in 3D TI thin films. Green area represents the TI thin film with its two surfaces (dark green), $t(k)$ the tunneling, and $2V$ an applied electrostatic potential. Yellow parts are $s$-wave superconductors, where the superconducting phase can be controlled by the magnetic flux $2\varphi$ (a) or an asymmetry can be induced in the intra-surface pairing strength (b). Depicted are both intra- (left) and inter-surface (right) Cooper pairs.
\label{fig:schem}}
\end{figure}

Figure \ref{fig:schem} shows two different schematics of the system we study. 
%
% FIGURE:
The TI thin film (green) is acting like a bilayer material with tunneling $t(k)$ between its two surfaces and is in proximity to a conventional superconductor (yellow). We also allow for an electrostatic potential difference $2V$ between the two surfaces, which causes a structure inversion asymmetry. The flux $\varphi$ in Fig.~\ref{fig:schem}(a) induces a phase difference between the pairing amplitudes in the two surfaces, but  at $\varphi=\pi$ time-reversal symmetry is still preserved.\cite{Keselmman-2013} In the setup in Fig.~\ref{fig:schem}(b) the intra-surface pairing is clearly different between the two surfaces. Inter-surface pairing can be prominent in both setups, and particularly in (b).\cite{Parhizgar14}
The effective low-energy Hamiltonian of the 3D TI thin film can be written as:\cite{Lu-2010,Shen-2010}
% EQUATION:
\begin{align}\label{eq:H TI}
h(k)=  \hbar v_F(\textbf{k} \cdot \sigma)\otimes \tau_z +
 t(k) \sigma_0 \otimes \tau_x 
 +V \sigma_0 \otimes \tau_z-\mu \sigma_0 \otimes \tau_0.
\end{align}
Here $\textbf{k}=(k_x,k_y)$ with $|\textbf{k}| = k$ and $v_F$ is the Fermi velocity, while $\sigma$ and $\tau$ are Pauli matrices in spin and surface space, respectively.
%Since TI surface states have a Dirac-like spectrum, the Hamiltonian only includes Rashba-like spin-orbit terms, with opposite signs on the two surfaces, and no conventional quadratic dispersion.
The inter-surface tunneling $t(k)$ in TI thin films is of the form $t_0-t_1k^2$.\cite{Zhang-2010}
%Bi$_2$Se$_3$:
Values for the intrinsic parameters for thin films of the prototypical TI Bi$_2$Se$_3$ are given in Table \ref{Tab1}. 
% TABLE:
\begin{table}[h]
\centering
 \begin{tabular}{ | c | c | c | c |}
    \hline
     & $v_F$ ($10^5$ ms$^{-1}$) & $t_0$ (eV) & $t_1$ (eV\AA$^2$) \\ \hline
    2 QLs & $4.71$ & $0.125$ & $21.8$ \\ \hline
    3 QLs & $4.81$ & $0.069$ & $18.0$ \\
    \hline
    4 QLs & $4.48$ & $0.035$ & $10.0$ \\ \hline
    \end{tabular}
    \caption{Parameters in the Hamiltonian Eq.~\eqref{eq:H TI} for Bi$_2$Se$_3$ TI thin films, extracted by fitting the dispersion relation with ARPES data (adopted from Ref.~\cite{Zhang-2010}). \label{Tab1} }
\end{table}
Due to a finite $t_0$, intrinsic Bi$_2$Se$_3$ thin films have a completely gapped spectrum.
Finally, tunable parameters are both the electrostatic potential $V$, by applying an electric field as a gate voltage, and the chemical potential $\mu$, which is adjustable by e.g.~chemical doping. In order to achieve a sustainable superconducting state we assume a finite doping, such that the Fermi level is located inside the conduction band.

\begin{figure}[h]
\centering
\includegraphics[width=0.4\linewidth]{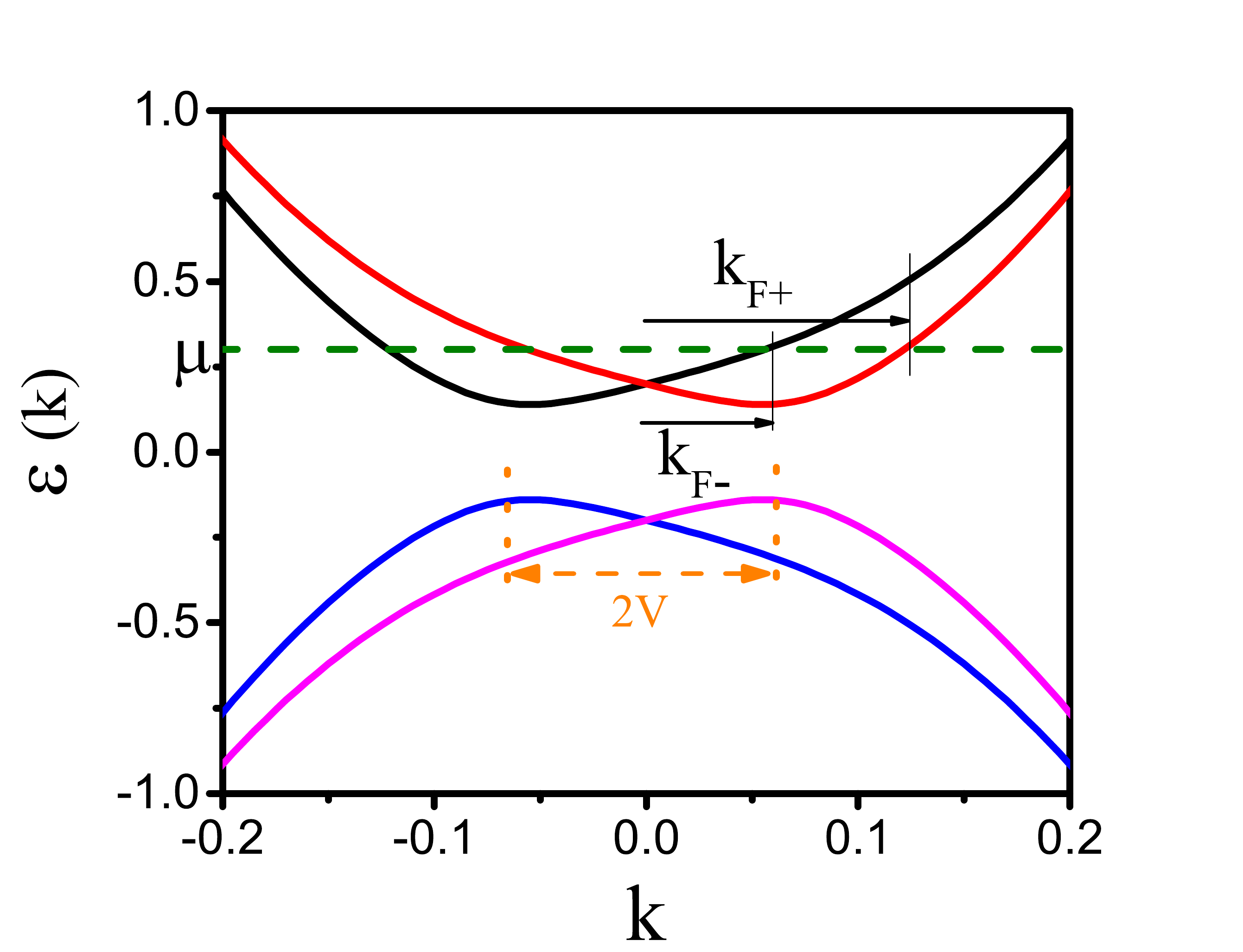}
\caption{Schematic figure of the band dispersion of a TI thin film in the normal state. Due to finite tunneling between the two surfaces, a gap occurs in the band dispersion. The potential $V$ gives rise to a Rashba splitting between the bands, such that at a finite chemical potential two Fermi circles with radii $k_{F-}$ and $k_{F+}$ appears.
\label{fig:disp}}
\end{figure}

The normal state band dispersion of the Hamiltonian in Eq.~\eqref{eq:H TI} is given by $\varepsilon(k)=\pm\sqrt{t(k)^2+(k\pm V)^2}$. By equating this energy to the chemical potential we find two different Fermi wavevectors $k_{F\pm}$, which produce two concentric Fermi level circles centered around the Brillouin zone center $k = 0$. Figure \ref{fig:disp} shows a schematic of the band dispersion of a TI thin film. As seen, a finite potential $V$ causes a Rashba splitting of the bands, which is the origin of the existence of the two Fermi circles with radii $k_{F\pm}$. Note that for all small to moderately large $\mu$ and $V$, all the relevant low-energy physics will be closely centered around the Brillouin zone center.

% SC:
Superconductivity in TIs can be either intrinsic\cite{intrinsic-SC1, intrinsic-SC2, intrinsic-SC3, intrinsic-SC4, intrinsic-SC5, intrinsic-SC6, intrinsic-SC7, FuBerg-2010, Fu-2014} or induced by proximity to an external superconductor.\cite{Wang-2012, Xu-2014} To allow for a general situation, we consider both Cooper pairs formed within each surface, intra-surface pairing $\Delta_{1}$ and $\Delta_2$, and between the two surfaces, inter-surface pairing $\Delta_{12}$. Note that inter-surface pairing can be quite prominent in setups such as Fig.~\ref{fig:schem}(b).\cite{Parhizgar14} Since higher momentum states are much more fragile in the presence of any disorder in systems with time-reversal symmetry according to Anderson's theorem \cite{Anderson}, we focus only on $s$-wave superconducting symmetries even when studying clean systems. It is also worth mentioning that an $s$-wave nature of the intrinsic superconductivity in Cu$_x$Bi$_2$Se$_3$ has been confirmed experimentally.\cite{Levy-2013}
This results in the total order parameter in the surface basis being
\begin{align}\label{eq:HSC}
\hat{\Delta}=\begin{pmatrix}
\Delta_1&\Delta_{12}^s+\Delta_{12}^t\\ \Delta_{12}^s-\Delta_{12}^t&\Delta_2
\end{pmatrix},
\end{align}
where the intra-surface pairing is necessarily spin-singlet, while inter-surface pairing can be either spin-singlet (s) or triplet (t), depending on the parity in surface index being even or odd, respectively. 
Finally, the total BdG Hamiltonian of the system, written in Nambu space, ($\psi^\dagger_k\,\psi_{-k}^T)$, is
% EQUATION:
\begin{equation}\label{H Nambu}
H(k)=\begin{pmatrix}
h(k)&\hat{\Delta}\\ \hat{\Delta}^\dagger & -h(-k)^T
\end{pmatrix}
\rm{\ \ \ where \ \ \ } \psi^\dagger_{k}=(c^\dagger_{k,1\uparrow}\,c^\dagger_{k,1\downarrow}\,c^\dagger_{k,2\uparrow}\,c^\dagger_{k,2\downarrow}).
\end{equation}

% ----------------------------------------- %
% TOPOLOGICAL ORDER
% ----------------------------------------- %
\section*{Results}
\subsection*{Topological order}
For a time-reversal-invariant system we can rotate the BdG Hamiltonian in Eq.~\eqref{H Nambu} and write it in the basis of
\begin{align}
\Psi_k=\frac{1}{\sqrt{2}}\begin{pmatrix}
\psi_k-i\hat{T}(\psi^\dagger_{-k})^T\\
\psi_k+i\hat{T}(\psi^\dagger_{-k})^T
\end{pmatrix}.
\end{align}
This results in
% EQUATION:
\begin{equation}
H(k)=\begin{pmatrix}
0&h(k)+i\hat{T}\Delta^\dagger\\ h(k)-i\hat{T}\Delta^\dagger&0
\end{pmatrix},
\end{equation} 
where $\hat{T} = \tau_0\otimes i\sigma_y \kappa$ is the time-reversal matrix, with $\kappa$ the complex conjugation operator. Following Ref.~\cite{Zhang} we now define the matrix element
% EQUATION:
\begin{equation}\label{delta s def}
\delta_n(k)=\langle \psi_n(k)\vert \hat{T}\Delta^\dagger \vert \psi_n(k)\rangle,
\end{equation}
where $\vert\psi_n(k)\rangle$ is the wave function of the $n$th band of the normal state Hamiltonian $h(k)$. 
In the weak pairing limit it is only the electrons very near the Fermi surface that are important for superconductivity, and the above equation can then be calculated at the two Fermi wavevectors $k_{F\pm}$ of the TI thin film.
Applying this expression in 2D, the topological order can be classified by the expression $N_{\rm 2D}=\Pi_n({\rm sgn}(\delta_n))^{m_n}$, where $m_n$ is the number of time-reversal-invariant points enclosed by the normal state Fermi surface(s). Since the only time-reversal symmetric point enclosed by the two Fermi circles is $k=(0,0)$, we find $m_n=1$.
Staying within the weak-pairing limit, $\delta_n(k)$ is only non-zero close to the Fermi surface and
for a fully-gapped superconductor it remains finite on all Fermi surfaces, which makes the sign of $\delta_n(k)$ a good topological number.
 After some calculations we also obtain the normalized wave functions corresponding to the two bands crossing the Fermi surface as
% EQUATION:
\begin{align}\label{eq:psi}
\vert \psi_n\rangle^T \! =\! \frac{1}{\sqrt{2}} \! \begin{pmatrix} \cos(\frac{\theta_n}{2}), n\cos(\frac{\theta_n}{2}) e^{i\phi_k}, \sin(\frac{\theta_n}{2}), n\sin(\frac{\theta_n}{2})e^{i\phi_k}\end{pmatrix}, \nonumber
\end{align}
where $n=\pm 1$ indicates the two different conduction bands and $\phi_k$ is the polar angle of the momentum. We have also defined the new variable $\theta_n$, such that $\sin(\theta_n)=t(k_{Fn})/\sqrt{(k_{Fn}+nV)^2+t(k_{Fn})^2}$.
Using the above wave functions, together with Eqs.~\eqref{eq:HSC} and~\eqref{delta s def}, we arrive at
% EQUATION:
\begin{equation}\label{delta s}
\delta_n=\frac{\cos^2(\frac{\theta_n}{2})}{2} [\Delta_1+\Delta_2\tan^2(\frac{\theta_n}{2})+2\Delta^s_{12}\tan(\frac{\theta_n}{2})],
\end{equation} 
for a spin-singlet $s$-wave superconducting state in a TI thin film. Note however that the spin-triplet inter-surface pairing is absent in this expression since it is odd under the surface index and thus necessarily gives a zero contribution in Eq.~\eqref{delta s def}.

% ----------------------------------------- %
% INTRA-SURFACE
% ----------------------------------------- %

\subsection*{Intra-surface pairing}
We first consider the case of zero or negligible inter-surface pairing and focus on only finite intra-surface orders $\Delta_i$. We assume that $\Delta_i$ is either produced by an intrinsic mechanism or by proximity effect to a conventional superconductor, in both cases resulting in spin-singlet $s$-wave symmetry.
The setup in Fig.~\ref{fig:schem}(a) for proximity-induced superconductivity allows for separately tuning the superconducting phases induced in the two surfaces of the TI thin film by controlling the magnetic flux. The two order parameter amplitudes can also be different by modifying the coupling between the TI thin film and the superconductor.

From Eq.~\eqref{delta s} it is clear that if the superconducting orders have the same sign, $\Delta_1 \Delta_2>0$, then $\delta_n$ always has the same sign for all bands. Thus, there exists no nontrivial topological phase if there is no phase difference between the superconducting states in the two surfaces. Limiting ourselves to time-reversal-invariant superconductivity, a $\pi$-shift is also allowed between the two superconducting states. This can clearly be achievable by tuning the flux in Fig.~\ref{fig:schem}(a). Alternatively, as has recently been found in 1D systems,\cite{Arbel1,Arbel2} possibly also driven by electron-electron interactions. In this latter case mean-field theory as well as density matrix renormalization group calculations were used to show that repulsive electron-electron interaction can result in a $\pi$-phase shift between two parallell systems.
Moreover, intrinsically $\pi$-shifted pairing terms have also recently been studied in general two-band systems. \cite{Deng-2012} Taken together, it is therefore relevant to consider this case also for TI thin films.
For the situation $\Delta_1\Delta_2<0$, the quantity $\tan^2(\theta_n/2)$ must be smaller than  $(-\Delta_1/ \Delta_2)$ on one of the Fermi surfaces but larger than it on the other in order to realize nontrivial topology. This means that the topological phase occurs between the two lines  $\Delta_1=-\tan^2(\theta_{\pm}/2)\Delta_2$. 

\begin{figure}[htb]
\centering
\includegraphics[width=0.99\linewidth]{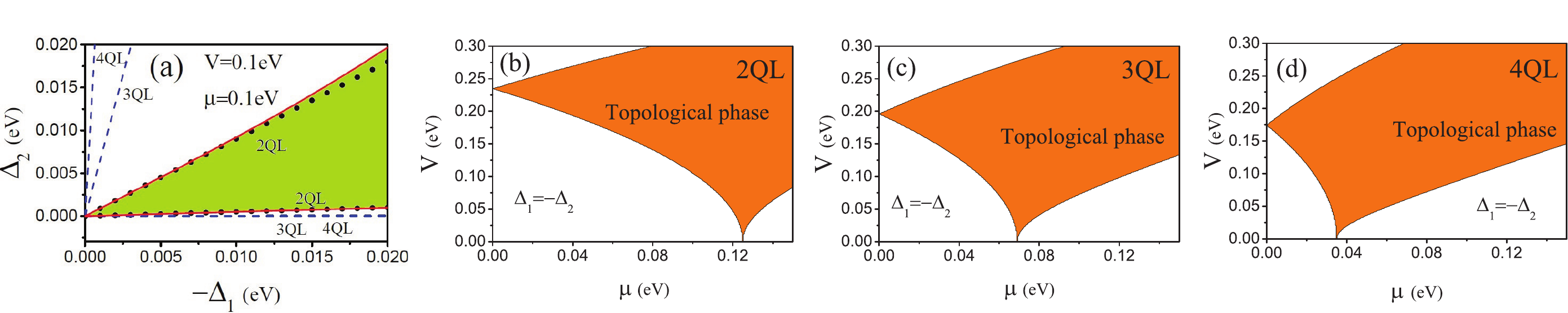}
\caption{Topological phase diagram for Bi$_2$Se$_3$ TI thin films for finite intra-surface pairing in the plane of (a) $(-\Delta_1,\Delta_2)$ and (b-d) $(V,\mu)$. In (a) dashed (3, 4 QLs) and dotted (2 QLs) lines display the topological criterion based on $\delta_n$ in Eq.~\eqref{delta s}, while solid red lines show the gap closing for the full Hamiltonian Eq.~\eqref{H Nambu} for 2 QLs. Here $\mu=0.1$~eV and $V=0.1$~eV. In (b-d) $\Delta_1 = -\Delta_2$ for 2-4 QL thick films with nontrivial phase in orange.   
\label{fig1}}
\end{figure}

Figure \ref{fig1}(a) shows the nontrivial topological phase for thin films of Bi$_2$Se$_3$ in the plane of $(-\Delta_1, \Delta_2)$ for $V=\mu=0.1$eV in the absence of any inter-surface pairing. Dashed/dotted lines show the condition $\Delta_1=-\Delta_2\tan^2(\theta_{\pm}/2)$ for thin films with thicknesses 2-4 QLs. Note how the topological phase grows with increasing film thickness. 
% FIGURE:
At the topological phase transition between the trivial and nontrivial phases the quasiparticle energy spectrum necessarily closes. The solid red lines in Fig.~\ref{fig1}(a) show the energy gap closing for the Hamiltonian Eq.~\eqref{H Nambu} for a 2 QL Bi$_2$Se$_3$ thin film, with the green area marking the topological nontrivial phase. The gap closing exactly coincides with the boundary of the topological region determined by the signs of $\delta_n$ for small values of $\Delta_i$, but there are small deviations for stronger superconductivity. This is not surprising since the topological criterion based on $\delta_n$ is strictly speaking only valid in the weak-pairing limit as it uses the normal state wave functions.
In Fig.~\ref{fig1} (b-d)  we investigate how the topological region (orange) varies with the tunable parameters of a TI thin film, the chemical potential $\mu$ and electric potential $V$, for a $\pi$-phase difference between the two superconductors proximitized to the TI thin film. Especially note that these results are not dependent on the amplitude of the $\Delta_i$'s.
As seen, a large part of the phase diagram is in the nontrivial topological phase for 2-4 QLs films. The nontrivial phase is even reachable for $\mu =0$ or $V = 0$, although much less fine-tuning is needed when applying an electric field to films with finite doping. This shows that it is easy to both achieve a time-reversal-invariant nontrivial topological state in TI thin films, and, most importantly, the topology is highly tunable by applying an electric field. 

\begin{figure}[htb]
\centering
\includegraphics[width=0.7\linewidth]{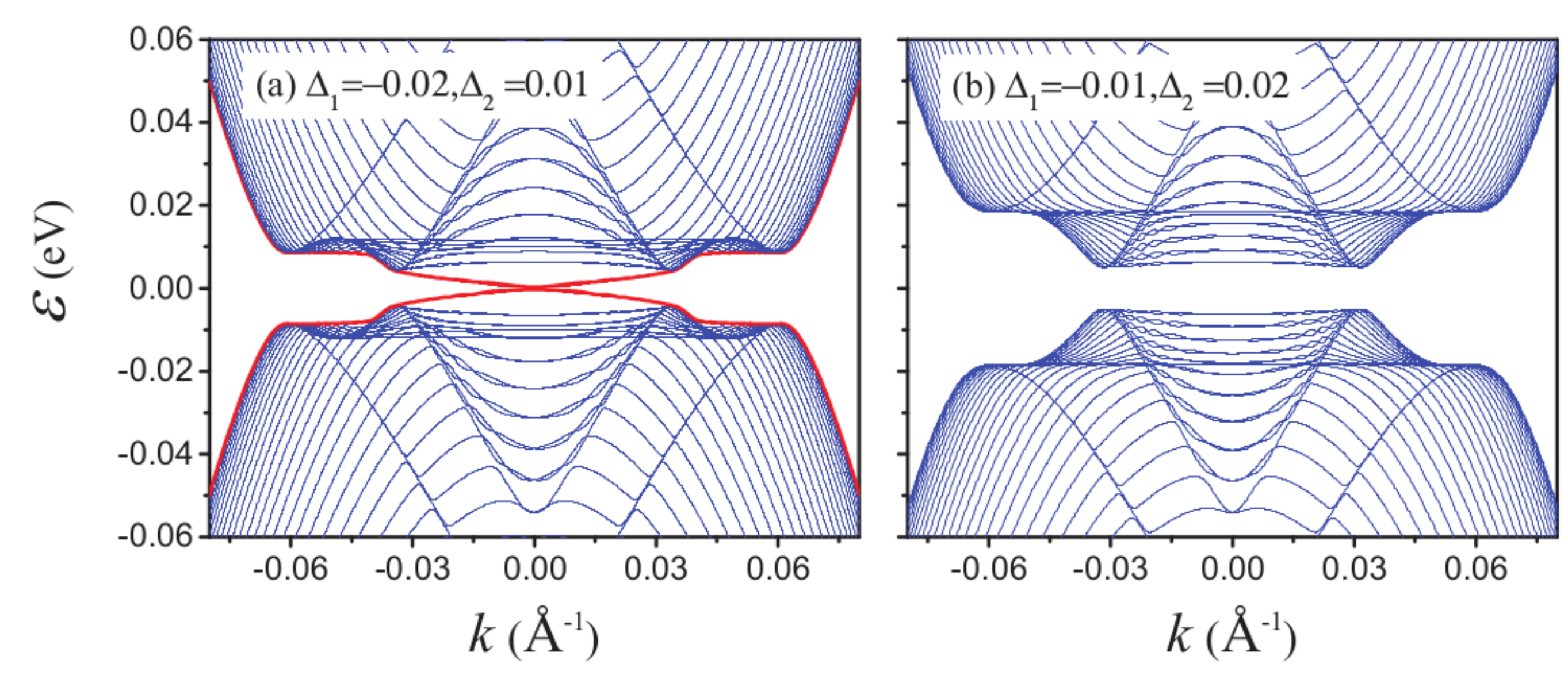}
\caption{Band dispersion of a bilayer Rashba nanoribbon, representing the Hamiltonian Eq.~\eqref{H Nambu} of a 2 QL Bi$_2$Se$_3$ thin film for $V=\mu=0.1$~eV and (a) $\Delta_1=-0.02$~eV, $\Delta_2=0.01$~eV, (b) $\Delta_1=-0.01$~eV, $\Delta_2=0.02$~eV with bulk states (blue) and edge states (red).
\label{fig3}}
\end{figure}

To illustrate that the nontrivial topological state hosts edge states we also calculate the band dispersion for an infinite (in the direction perpendicular to the edge) nanoribbon of a bilayer Rashba material on a square lattice with parameters such that the normal state Hamiltonian represents Eq.~\eqref{eq:H TI} for a 2 QL Bi$_2$Se$_3$ TI thin film. 
In Fig.~\ref{fig3}(a) we plot the band dispersion for $\Delta_1=-0.02$~eV, $\Delta_2=0.01$~eV and $V=\mu=0.1$~eV, which is well within the topological phase according to Fig.~\ref{fig1}(a). There is a Kramers pair of Majorana modes (red lines) crossing the bulk superconducting gap, which are very well localized to the edges of the nanoribbon. This confirms the nontrivial topology of the system. In Fig.~\ref{fig3}(b) we instead plot the band dispersion for $\Delta_1=-0.01$~eV and $\Delta_2=0.02$~eV where the topology is trivial according to Fig.~\ref{fig1}(a) and confirmed by the fully gapped spectrum.
% FIGURE:
%

% ----------------------------------------- %
% INTER-SURFACE:
% ----------------------------------------- %
\subsection*{Inter-surface spin-singlet pairing}
Next we also allow for inter-surface spin-singlet pairing. Intrinsic inter-surface pairing should always be considered in systems where the surfaces are close to each other. This applies to both intrinsic and proximity-induced superconductivity. As we show below, ultrathin films of Cu-doped Bi$_2$Se$_3$ have strong inter-surface pairing if the bulk is in one of the currently proposed inter-orbital pairing states.\cite{FuBerg-2010, intrinsic-SC1,intrinsic-SC3,intrinsic-SC4, Fu-2014} In proximity-induced systems inter-surface pairing can also be strong, especially when generated by crossed Andreev reflection.\cite{CAR1, CAR2} In this process the Cooper pair is split between the two surfaces, which is generally possible when the superconducting penetration depth is larger than the film thickness.
Here we restrict the discussion to the same sign for $\Delta_1,\Delta_2$, and $\Delta_{12}^s$, which is the simplest proximity-induced setup with no additional fluxes. [We also find that a $\pi$-shift in the intra-surface pairing combined with finite inter-surface pairing gives large nontrivial topological regions, as well as other combinations where a subset of the order parameters are $\pi$-shifted.]
Under this condition, the sign of $\tan(\theta_n/2) = [\sqrt{(k_{Fn}+nV)^2+t(k_{Fn})^2}-(k_{Fn}+nV)]/t(k_{Fn})$ is decisive in determining the sign of $\delta_n$ in Eq.~\eqref{delta s}. As a direct consequence, nontrivial topology is only possible if $t(k)$ is negative for at least one band. In addition, for positive and equal intra-surface pairing $\Delta_1=\Delta_2$, as expected if the superconductor couples equally well to both surfaces, as in Fig.~\ref{fig:schem}(a), the inter-surface pairing also needs to fullfil $\Delta^s_{12}>\Delta_1=\Delta_2$ in order to reach a nontrivial state. Near the limit $\Delta_i = 0$, the sign of $\delta_n$ is simply equal to the sign of $t(k_{Fn})$ on each Fermi surface. Such dominating inter-surface pairing can easily be a consequence of intrinsic pairing. It has also been found for proximity-induced superconductivity in double 1D wire systems with strong electron-electron interactions, where the repulsive interaction heavily suppresses intra-wire pairing, while inter-wire pairing is significantly less affected.\cite{Gangadharaiah-2011,Stoudenmire-2011,Recher-2002,Flensberg-2014, Jelena-2014, Danon2015, Ebisu2015} Also in 2D layer systems, dominating inter-layer pairing has recently been studied.\cite{Nakosai-2012, Hosseini} In addition, it has recently been found that screening has a stronger effect on intra-layer properties than on inter-layer ones in double layer graphene systems.\cite{Fariborz-magres} Together these results clearly validates studying also the strong inter-surface pairing regime.
Note though that we do not explicitly include the effects of electron-electron interactions. However, since the derived  conditions for non-trivial topology are independent on the size of the individual pairing amplitudes, any band or pairing renormalizations arising from interaction effects are actually automatically included. Thus, it is possible to study also this regime within our framework.

\begin{figure}[htb]
\centering
\includegraphics[width=0.99\linewidth]{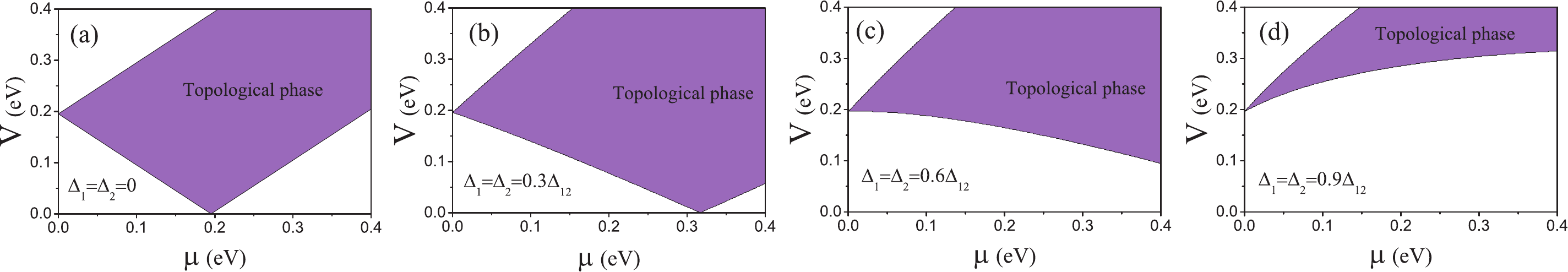}
\caption{Topological phase diagram for a 3 QL Bi$_2$Se$_3$ TI thin film for finite spin-singlet inter-surface pairing in the plane of $(V,\mu)$ when (a) $\Delta_1=\Delta_2=0$, (b) $\Delta_1 \Delta_2=0.3\Delta^s_{12}$, (c) $\Delta_1=\Delta_2=0.6\Delta^s_{12}$, and (d)  $\Delta_1=\Delta_2=0.9\Delta^s_{12}$.
\label{fig4}}
\end{figure}

In Fig.~\ref{fig4} we plot the topological phase of a 3 QL Bi$_2$Se$_3$ thin film for finite inter-surface pairing with different intra-surface contributions as function of the two tunable parameters $V$ and $\mu$ when $\Delta_1 = \Delta_2$. This setup is expected for a superconductor coupling equally well to both surfaces.
As the intra-surface pairing approach the inter-surface pairing, the nontrivial topological region (purple) shrinks, but a large nontrivial region is present even for substantially large intra-surface pairing amplitudes. Note that these results are not dependent on the actual amplitudes of the different superconducting order parameters, but only on their ratios.
%FIGURE:

\begin{figure}[htb]
\centering
\includegraphics[width=0.5\linewidth]{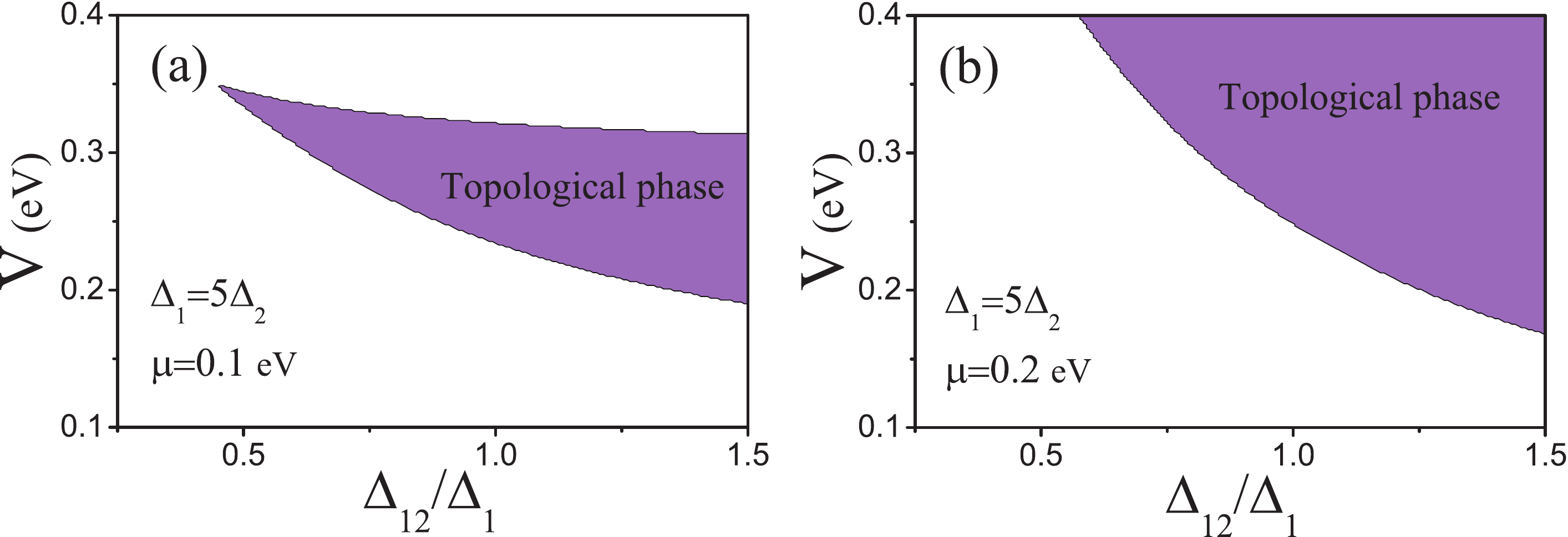}
\caption{Topological phase diagram for a 3 QL Bi$_2$Se$_3$ TI thin film for chemical potentials (a) $\mu=0.1$~eV and (b) $\mu=0.2$~eV in the plane of $(V,\Delta_{12}^s)$ when the intra-surface pairing is highly assymmetric, here $\Delta_1=5\Delta_2$.
\label{fig5}}
\end{figure}

Alternatively to Fig.~\ref{fig4} we also consider a superconductor proximity-coupled to only the first TI surface, as in Fig.~\ref{fig:schem}(b). In the ultrathin limit, earlier results have found both finite intra-surface pairing in both surfaces, albeit $\Delta_1\gg \Delta_2$, and also significant inter-surface pairing.\cite{Parhizgar14} In Fig.~\ref{fig5} we show the topological phase diagram in the plane of $V$ and $\Delta_{12}^s/\Delta_1$ for a fixed typical value of $\Delta_1=5\Delta_2$ and two different chemical potentials. The results clearly show that a non-trivial phase is possible for inter-surface order parameter $\Delta_{12}^s$ notably smaller than $\Delta_1$, and that again, the results is highly tunable with $V$ and $\mu$.

The clear dependence on the size of the inter-surface pairing is somewhat reminiscent to a 1D double Rashba wire system, where topological superconductivity has been found for dominating inter-wire pairing.\cite{Flensberg-2014} However, note that the topology of TI thin films is not only dependent to this ratio of the pairing amplitudes, but also very sensitive to the sign of the tunneling function $t(k)$ on each Fermi surface. Since this $t(k_{Fn})$ can very readily be tuned by both chemical doping $\mu$ and electric potential $V$, it leads to topology being easily controllable for TI thin films, in contrast to the 1D systems studied previously.

\subsection*{Inter-surface spin-triplet pairing}
Finally we also investigate spin-triplet inter-surface pairing. A finite spin-triplet inter-surface pairing $\Delta_{12}^t$ always gives zero contribution to Eq.~\eqref{delta s def} since it is odd under the surface index, and can thus not be investigated within the above framework. Instead, we make use of previous work on the possible topological phases in the bulk superconducting state in Cu-doped Bi$_2$Se$_3$, which is one of the most studied superconducting TIs.\cite{intrinsic-SC1, intrinsic-SC2, intrinsic-SC3, intrinsic-SC4, intrinsic-SC5, intrinsic-SC6, intrinsic-SC7, FuBerg-2010, Fu-2014}
Using the lattice symmetries of the centrosymmetric bulk Bi$_2$Se$_3$, the superconducting pair amplitudes in Cu$_x$Bi$_2$Se$_3$ have been analyzed in terms of their orbital structure, divided into intra-orbital spin-singlet and inter-orbital spin-singlet and triplet states.\cite{FuBerg-2010, Fu-2014} It was shown that a fully gapped time-reversal-invariant topological superconducting state always exists in this centrosymmetric material for an odd inter-orbital order parameter, which is necessarily in a spin-triplet state.\cite{FuBerg-2010}

For the purpose of thin films, we first assume that the thin film limit of Cu$_x$Bi$_2$Se$_3$ is given by the Hamiltonian Eq.~\eqref{eq:H TI} with only a doping-induced shift of the chemical potential. This is very reasonable since the bulk band dispersion has experimentally been shown to not vary significantly with Cu-doping.\cite{intrinsic-SC1, intrinsic-SC2}
Moreover, the upper surface of a TI thin film is almost exclusively composed of one orbital, while the states on the opposite surface come from the other orbital.\cite{Yip-2013} Thus we can generalize the bulk results based on orbital parity\cite{FuBerg-2010} and apply them to the surface parity of a TI thin film. Using this line of argument, we directly find a fully gapped nontrivial topological phase for odd inter-surface spin-triplet superconducting pairing $\Delta_{12}^t$ in a TI thin film. Importantly, this directly implicates that thin films of Cu-doped Bi$_2$Se$_3$ will have strong dominating spin-triplet inter-surface pairing, since recent results strongly suggests that the bulk is in an odd inter-orbital pairing state.\cite{Fu-2014, intrinsic-SC6}
Ultrathin films of Cu-doped Bi$_2$Se$_3$, or similarly intrinsic superconducting TIs, are thus a very strong candidate for inter-surface pairing forming a 2D topological state with a Kramers pair of helical Majorana 1D edge states. 
This state is not tunable solely by chemical potential, but applying an electrostatic potential $V$ breaks the structure inversion symmetry, which technically invalidates the above proof of a nontrivial topological state. By numerically studying the energy dispersion of TI thin film nano-ribbons, we find that the topologically protected zero-energy edge states actually prevail to moderately high $V$, but that the topology becomes trivial above a critical $V$.

% ----------------------------------------- %
% SUMMARY
% ----------------------------------------- %
\section*{Conclusion}
We have investigated the possibilities for time-reversal-invariant topological superconductivity in 3D TI thin films. For pure intra-surface pairing we show that topological superconductivity requires the order parameters to have different signs on the two different surfaces. Within this regime, the topology is highly tunable by both doping and electric field. 
Including strong inter-surface pairing alters the criteria for nontrivial topology. For spin-singlet inter-surface pairing, nontrivial topology requires the inter-surface tunneling to change sign with momentum, a feature very special to TI thin films and highly tunable. Notably, when the intra-surface pairing is asymmetric, for example due to superconducting proximity effect primarily to only one surface, the nontrivial state is reached even for inter-surface pairing subdominant compared to the intra-surface pairing. 
For spin-triplet inter-surface pairing we prove that the superconducting state is always nontrivial, since it is necessarily odd in the surface index. We also show that the recently suggested odd inter-orbital pairing state in bulk Cu-doped Bi$_2$Se$_3$\cite{Fu-2014, intrinsic-SC6} results in dominating spin-triplet inter-surface pairing and thus a nontrivial topological state in TI thin films. 
These results show that multiple different topological states can exists in superconducting TI thin films. Moreover, the topology is most often highly dependent on the applied electric field and chemical potential, which opens for exciting prospects of {\it in-situ} tuning of time-reversal invariant topological superconductivity and its associated Kramers pair of Majorana edge modes in TI thin films.
These results do not only generalize earlier results for double 1D wires to the case of higher spatial dimensions, but importantly shows that the 3D TI thin films offers much more tunability and control of the topological phases.

\section*{Acknowledgements}

The authors thank A.~Bouhon, K.~Bj\"{o}rnson, and D.~Kuzmanovski for discussions. This work was supported by the Swedish Research Council (Vetenskapsr\aa det), the G\"{o}ran Gustafsson Foundation, the Swedish Foundation for Strategic Research (SSF), and the Wallenberg Academy Fellows program. 

\section*{Author contributions statement}

A.B.S. proposed the idea and F.P. performed the calculations. Both authors analyzed the results and wrote the article.

\section*{Additional information}
\textbf{Competing interests:} The authors declare no competing financial interests.

During the final stages of completion of this work, Ref.~\cite{Wang2016} appeared, which treats TI thin films but only includes intra-surface pairing. The results in this limit are similar to ours.

\end{document}